\documentclass{optica-article}

\journal{opticajournal} 

\articletype{Research Article}

\usepackage{lineno}

\usepackage{xcolor}
\usepackage{soul}

\usepackage{comment}

\begin{document}

\title{Quantum scanning synthetic optical holography}

\author{\author{Josué R. León-Torres,\authormark{1, 2, 3*} Byron Caiza,\authormark{1, 2} Nadia Baumann,\authormark{4} Sebastian Töpfer,\authormark{5} Anna Mühlig,\authormark{4} Orlando Guntinas-Lichius,\authormark{4} Karin Burger,\authormark{2} Frank Setzpfandt,\authormark{1, 2} Markus Gräfe,\authormark {2, 5} and Valerio Flavio Gili\authormark{2}}}

\address{
\authormark{1}Abbe Center of Photonics, Friedrich Schiller University Jena, Albert-Einstein-Straße 6, 07745 Jena, Germany\\

\authormark{2}Fraunhofer Institute for Applied Optics and Precision Engineering IOF, Albert-Einstein-Straße 7, 07745 Jena, Germany\\

\authormark{3}Cluster of Excellence Balance of the Microverse, Friedrich Schiller University Jena, Jena, Germany\\

\authormark{4}Universitätsklinikum Jena, Am Klinikum 1, 07747 Jena, Germany\\

\authormark{5}Institute for Applied Physics, Technical University of Darmstadt, Otto-Berndt-Straße 3, 64287 Darmstadt, Germany
}

\email{\authormark{*}josue.ricardo.leon.torres@iof.fraunhofer.de} 

\begin{abstract*} 
Synthetic optical holography (SOH) introduced holographic reconstruction into scanning optical microscopy, enabling quantitative phase imaging with sequential acquisition and point detection. Here, we extend this concept to the quantum regime by implementing SOH within quantum imaging with undetected light (QIUL). By integrating a controlled synthetic phase carrier into a scanning QIUL implementation, we retrieve amplitude and phase images of objects probed by mid-infrared (MIR) photons while detecting only their visible partners. We demonstrate the method on binary, transparent, and biological samples, showing complex-field reconstruction in a scanning quantum imaging system. This approach decouples spatial resolution from photon-pair spatial correlations and establishes a route toward diffraction-limited, label-free MIR phase imaging with visible-wavelength detection.
\end{abstract*}

\section{Introduction}

Scanning optical microscopy is widely used when flexible illumination geometries, point detection, or spatially localized probing are required. However, the sequential nature of image acquisition has historically made the direct implementation of holography challenging, limiting the ability of scanning schemes to retrieve quantitative phase information. Synthetic optical holography (SOH) was introduced to overcome this limitation by encoding a synthetic spatial carrier through controlled relative motion between the object and reference fields, thereby enabling holographic reconstruction in scanning microscopy without relying on camera-based wide-field detection \cite{synthetic_schnell}. Following this first demonstration, several adaptations of SOH for optical microscopy were developed, establishing it as a practical route toward quantitative phase imaging in scanning systems  \cite{Schnell2014SyntheticHolography, CanalesBenavides2019SinusoidalSOH, Deutsch2014NonlinearPhaseSOH}.

A similar challenge remains in quantum imaging with undetected light (QIUL). QIUL is a nonlinear-interferometric imaging technique based on photon pairs, typically generated at different wavelengths through spontaneous parametric down-conversion (SPDC). In this approach, one photon probes the object but is never detected, while its partner, which does not interact with the object, is measured \cite{Lemos2014, Jorge_Review_2024, ArminReview, Marta_Review_2019, Defienne2024}. The object information is transferred to the detected photon through induced coherence, which arises when the two possible photon-pair generation processes become indistinguishable \cite{Mandel91, Mandel1991_2}. This enables illumination at one wavelength while detecting at another, offering a powerful path toward mid-infrared (MIR) probing with visible-wavelength detection \cite{Inna2020, Placke26}.

Recent work has extended QIUL to scanning microscopy, demonstrating that point-wise acquisition can decouple the spatial resolution from the position or momentum correlations of the photon pairs \cite{Leon-Torres2025}. In such a scanning architecture, the spatial resolution is instead determined by the classical optical properties of the imaging system, and can in principle approach the diffraction limit set by the illumination optics. However, this previous implementation retrieved only intensity information and did not provide a mechanism for quantitative phase reconstruction. Thus, despite the progress in scanning QIUL, a method combining scanning-based quantum imaging with full complex-field retrieval has remained absent.

In parallel, several works have demonstrated holographic phase imaging within QIUL using wide-field configurations \cite{Sebastian2022, Leon-Torres2024, Topfer25, Leon-Torres2025_QD, Pearce2024, Pearce2023QIUL}. These approaches have shown that undetected-light interferometry can recover phase information, but they rely on camera-based acquisition and therefore do not address the specific constraints and opportunities of scanning architectures. In particular, a scanning holographic implementation of QIUL has not yet been demonstrated.

Here, we present the first implementation of scanning SOH within QIUL. By integrating a controlled synthetic phase carrier into a scanning QIUL scheme, we retrieve both amplitude and phase images of objects illuminated by MIR light while detecting their visible partners. We demonstrate the method using binary, transparent, and biological samples, showing complex-field reconstruction in a scanning quantum imaging system. This work establishes scanning synthetic holography as a route toward label-free MIR phase imaging with visible detection, while preserving the resolution advantages of point-scanning acquisition.

\section{Scanning synthetic optical holography via undetected light}
\label{sec:synthetic}
The implementation of holography spans a wide range of fields, with microscopy and optical imaging being among the most prominent examples. Several approaches, such as digital phase-shifting holography \cite{Yamaguchi1997, Schnars2015}, off-axis holography (OAH) \cite{Leith62, Kim_Holo, Sanchez}, and phase-retrieval algorithms such as the Gerchberg–Saxton method \cite{Gerchberg1972, Schnars2015}, are typically implemented in wide-field configurations to recover the phase information carried by an optical field. In scanning optical holography, however, the available holographic approaches are  limited, often requiring multi-frame acquisition schemes that increase the total imaging time.

A suitable alternative is SOH \cite{synthetic_schnell, Schnell2014SyntheticHolography, CanalesBenavides2019SinusoidalSOH, Deutsch2014NonlinearPhaseSOH}, which is conceptually equivalent to OAH but eliminates the need to introduce a relative angle between the object and reference beams. Instead, the required linear phase term is generated through controlled relative motion between the object and reference beams. This approach preserves single-shot phase retrieval while significantly simplifying the experimental implementation. In the following, we describe how SOH is integrated into the raster-scanning QIUL scheme introduced in \cite{Leon-Torres2025} to allow a full reconstruction of the complex transmission function of the object with undetected light.

\subsection{Working principle and experimental setup}

Unlike conventional multi-frame interferometric methods, where each image is acquired at a fixed reference mirror position, SOH encodes the phase information through a continuous displacement of the reference mirror during raster scanning. The resulting phase accumulation between the object and reference fields enables reconstruction of the complex object field from a single synthetic hologram \cite{synthetic_schnell, Schnell2014SyntheticHolography}.

\begin{figure}[ht!]
\centering\includegraphics[width=0.99\textwidth]{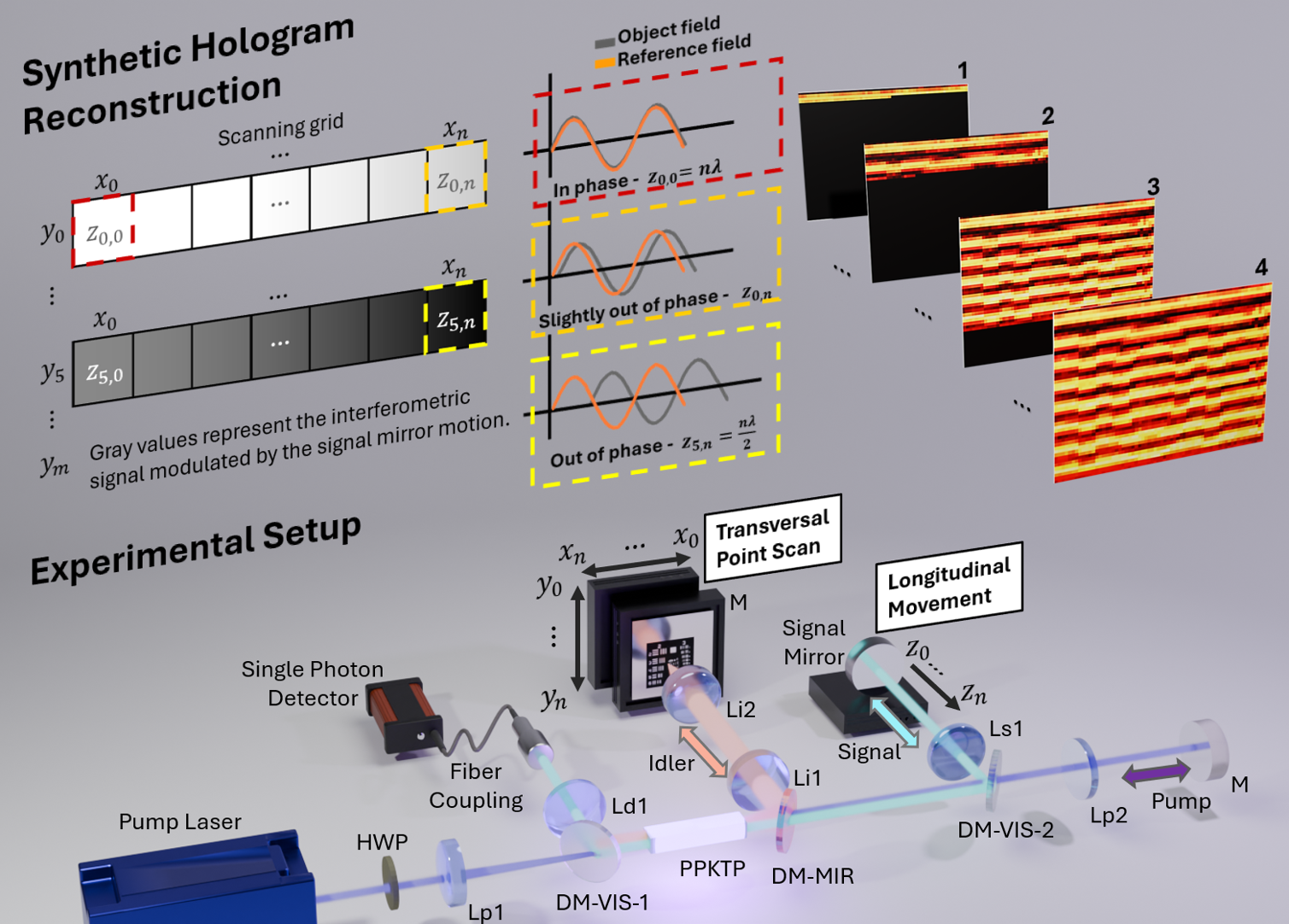}
\caption{Schematic illustration of the SOH implementation in the QIUL scanning microscope. Top: During raster scanning, the sample is translated transversely while the signal mirror is displaced longitudinally at each scan position to introduce a controlled phase shift between consecutive measurements. The accumulated phase is encoded as gray values in the scanning grid, from which a synthetic hologram is reconstructed after completion of the scan. Bottom: Experimental implementation of raster-scanning microscopy with undetected light integrated with SOH. The pump, signal, and idler beams are shown in purple, cyan, and red, respectively. The transverse raster scan of the sample and longitudinal displacement of the signal mirror jointly encode the phase-modulated intensity measurements to form the synthetic hologram.}
\label{fig:Fig_1}
\end{figure}

This process is illustrated in the top panel of Fig.~(\ref{fig:Fig_1}), which shows the scanning grid on the left. The gray value at each scan position represents the recorded intensity corresponding to the accumulated relative phase between the object and reference fields, introduced by the incremental displacement of the signal mirror.

As the scan progresses, the displacement of the signal mirror changes the optical path difference, causing the object and reference fields to alternate between constructive and destructive interference. This evolution is highlighted by dashed squares, where red denotes in-phase fields (fully constructive interference), orange indicates a partial phase mismatch, and yellow corresponds to out-of-phase fields (fully destructive interference). The total longitudinal displacement is determined by the number of scan positions along the $x$- and $y$-axes, denoted by $n$ and $m$, respectively, together with the step size $\Delta_z$, which results in a total displacement of $nm\Delta_z$. The resulting intensity modulation, illustrated on the right of the top panel, shows the sequential formation of the synthetic hologram. After completion of the scan, the discrete displacements of the signal mirror appear as the characteristic interference fringe pattern in the reconstructed hologram.

A schematic of the experimental setup for quantum scanning SOH via undetected light is shown in the lower panel of Fig.~(\ref{fig:Fig_1}), which highlights a Michelson interferometer–type configuration, where the photon pairs, conventionally referred to as signal and idler, are generated via SPDC \cite{Burnham1970, Mandel1985_Th}. A continuous-wave laser at 405\ nm (shown in purple) bidirectionally pumps a 30\ mm long ppKTP crystal with 60~mW of optical power. The forward and backward generated signal beams at 487\ nm propagate along the paths indicated in cyan, while the corresponding idler beams at 2400\ nm follow the optical paths shown in red. An ultra-narrow band-pass filter with a bandwidth of 1.8~nm is placed on the signal path to suppress background contributions. 

The forward-generated idler field is focused down to $27.8 \pm 5\ \mu\mathrm{m}$ onto the sample surface by lenses $L_{\mathrm{i1}}$ and $L_{\mathrm{i2}}$ \cite{Leon-Torres2025}. The light scattered by the sample is subsequently directed back to the crystal plane, where it induces coherence between the two photon-pair generation processes. 


Building upon our previous work \cite{Leon-Torres2025}, the implementation of SOH requires the introduction of a controlled longitudinal phase modulation within the scanning QIUL system. Experimentally, this is achieved through a slow and continuous translation of a linear piezo stage that controls the optical path of the forward-generated signal field. To ensure that the induced displacement modifies only the optical phase without introducing significant alignment variations, the signal beam is configured in a 2-$f$ imaging system, which collimates the beam along its path to the signal mirror.

In QIUL, image formation arises from the interference of the probability amplitudes associated with the forward- and backward-generated signal beams reaching the detection plane. \cite{Feynman1951, Mandel1991CoherenceIndistinguishability, Glauber_NP}. This interference is enabled by the indistinguishability of the two generation processes, which is established through induced coherence when the optical paths of the idler beams are properly aligned \cite{Mandel91, Mandel1991_2}.

The critical distinction between classical and quantum implementations of SOH lies in the physical interpretation of the object and reference fields \cite{Leon-Torres2024, Topfer25, Leon-Torres2025_QD}. In the present configuration, the forward-generated idler beam is focused onto the object plane, while the sample is raster-scanned beneath a fixed illumination spot. Although the idler beams are never detected, the object information is transferred to their signal counterparts through induced coherence. Consequently, the forward-generated signal field carries the object information and acts as the object field. This beam is directed toward a mirror mounted on a linear translation stage, introducing a slowly varying phase through precise control of its optical path. In contrast, the backward-generated signal beam, whose phase remains constant throughout the scanning process, serves as the reference field, providing a spatially uniform phase. The superposition of these fields gives rise to a synthetic hologram, characterized by interference fringes determined by the virtual wave vector $\boldsymbol{k}_{\parallel}= (k_x, k_y)$ \cite{synthetic_schnell, Schnell2014SyntheticHolography}, with
\begin{equation}
k_x = \frac{4\pi}{\lambda_s}\,\frac{v_s}{v_x},
\qquad
k_y = \frac{4\pi}{\lambda_s}\,\frac{v_s}{\Delta_y\, v_x / (2X)}.
\end{equation}

Here, $\lambda_s$ denotes the wavelength of the detected signal field to which the longitudinal phase ramp is applied. $v_{\mathrm{s}}=\frac{\Delta_z}{t_\mathrm{step}}$ is the  velocity of the signal mirror with $t_{\mathrm{step}} = \left(t_{\mathrm{exe}} + t_{\mathrm{exp}} + t_{\mathrm{sta}}\right)$,
where \(t_{\mathrm{exe}}\) is the execution time, \(t_{\mathrm{exp}}\) is the detector exposure time, and \(t_{\mathrm{sta}}\) is the stabilization time between consecutive scan positions. Unless stated otherwise, all measurements were acquired with  \(t_{\mathrm{exp}}=0.25~\mathrm{s}\) and \(t_{\mathrm{exe}} + t_{\mathrm{sta}}\sim 0.50~\mathrm{s}\). $v_x=\frac{\Delta_x}{t_\mathrm{step}}$ is the scan velocity in the x-axis, and $X= n \Delta_x$ is the length of each scanned row, with $\Delta_x$ denoting the step size and $n$ the number of steps. $\Delta_y$ represents the step size in the y-axis, which is applied after the completion of each fast-scan line along the $x$-axis. Throughout this work, $\Delta_y =\Delta_x$. For the parameters used in these measurements, the transverse scanning velocity \(v_x\) ranges from \(6.7~\mathrm{\mu m/s}\) to \(67~\mathrm{\mu m/s}\), while the longitudinal signal mirror velocity \(v_s\) ranges from \(0.3 \pm 4.1~\mathrm{nm/s}\) to \(1.5 \pm 3.3~\mathrm{nm/s}\) (mean \(\pm\) standard deviation), reflecting the step-to-step fluctuations of the piezo motion.

The resulting synthetic field is equivalent to the tilted reference wave employed in wide-field OAH. Consequently, the amplitude and phase of the object field can be recovered using standard holographic reconstruction methods \cite{Verrier, Sanchez, Cuche}. 

The resulting quantum synthetic hologram is given by
\begin{equation}
I(\mathbf{r}) \sim 1 + |T(\mathbf{r})|^2 \cos \left[ \Delta\phi + \phi_{\text{synthetic}}(\mathbf{r}) + 2\phi_{\text{object}}(\mathbf{r}) \right],
\label{eq:1}
\end{equation}

where $T(\mathbf{r})=|T(\mathbf{r})|e^{i\phi_\mathrm{object}(\mathbf{r})}$ is the complex transmission function of the object, with $|T(\mathbf{r})|$ denoting the amplitude transmission and $\phi_{\text{object}}(\mathbf{r})$ the phase imparted by the object. $\Delta\phi$ is the overall interferometric phase, and $\phi_\text{synthetic}(\mathbf{r})=\mathbf{k}_{\parallel} \cdot \mathbf{r}$ is the phase introduced by the relative motion of the stages. The square factor and the factor of two account for the double pass through the object. 

The resulting synthetic hologram given by Eq.~(\ref{eq:1}) exhibits an interference fringe pattern with spatial period $\frac{2\pi}{|\mathbf{k}_{\parallel}|}$, which shifts the different contributions of the recorded intensity in Fourier space, enabling their separation during reconstruction. Appropriate control of the virtual wave vector $\boldsymbol{k}_{\parallel}$ ensures an optimal separation of these contributions, allowing the object information to be isolated and the complex object field to be recovered.

A Fourier transform of Eq.~(\ref{eq:1}) reveals the three distinct terms of the intensity distribution in the Fourier space, which is expressed as
\begin{equation}
\tilde{I}(\mathbf{q}) \sim 2\pi\ \delta(\mathbf{q}) + e^{i\Delta\phi} \mathcal{F}\left\{ |T(\mathbf{r})|^2 e^{i2\phi_{\text{object}}(\mathbf{r})} \right\}(\mathbf{q} - \mathbf{k}_{\parallel}) + e^{-i\Delta\phi} \mathcal{F}\left\{ |T(\mathbf{r})|^2 e^{-i2\phi_{\text{object}}(\mathbf{r})} \right\}(\mathbf{q} + \mathbf{k}_{\parallel}).
\end{equation}

The first term corresponds to the zero-frequency (DC) component or the uniform background illumination. The operator $\mathcal{F}$ indicates the Fourier transform. The synthetic wave-vector $\mathbf{k}_{\parallel}$ acts as a spatial carrier frequency, shifting the object information into off-axis sidebands at $\mathbf{k}_{\parallel} \pm  \mathbf{q}$. These sidebands are commonly referred to as the direct and conjugate terms and can be isolated through Fourier-domain filtering.

\subsection{Image processing}
\begin{figure}[ht!]
\centering\includegraphics[width=0.98\textwidth]{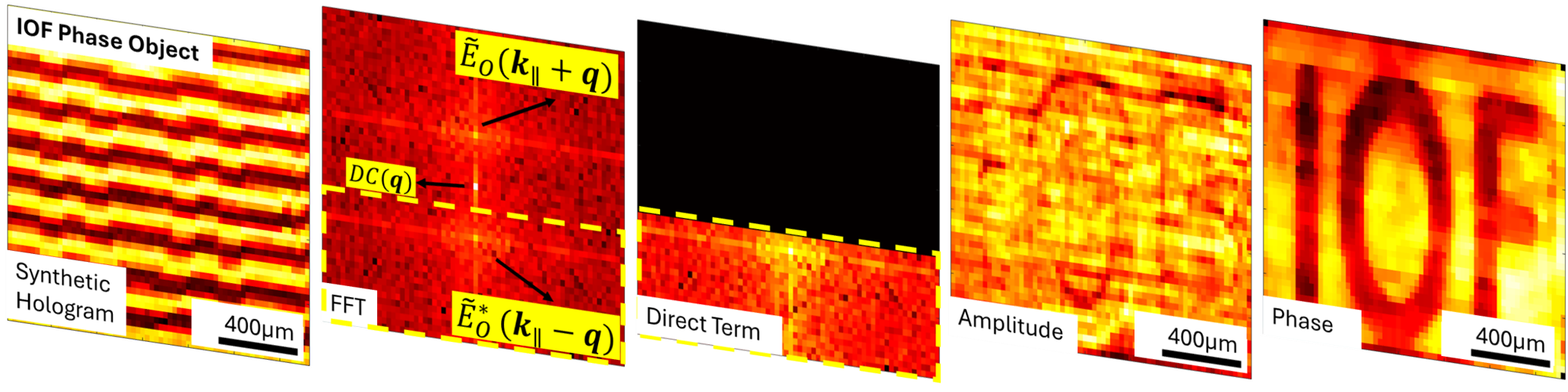}
\caption{Synthetic hologram reconstruction and complex-field retrieval. From left to right: synthetic hologram recorded during raster scanning, where slight distortions reveal the presence of the transparent object; corresponding Fourier spectrum showing the DC term and the spectrally shifted direct and conjugate terms; isolated direct term obtained by Fourier filtering; and the reconstructed amplitude and phase images of the object after inverse Fourier transformation, yielding a complex-valued field.} 
\label{fig:Fig_Ch4_11}
\end{figure}

The image reconstruction procedure employed in this work builds on the standard processing steps of classical OAH \cite{Sanchez, Schnell2014SyntheticHolography, Kim_Holo}, illustrated in Fig.~(\ref{fig:Fig_Ch4_11}). This process enables the retrieval of both amplitude and phase information of the object. Figure~(\ref{fig:Fig_Ch4_11}) presents the different reconstruction steps for a glass-plate sample with the letters “IOF” engraved on its surface. From left to right, the synthetic hologram exhibits horizontal fringe patterns generated by the interference between the object and reference beams. A slight vertical displacement of the fringes indicates the presence of the transparent sample. Applying a fast Fourier transform (FFT) reveals the characteristic holographic components: the DC, direct, and conjugate terms. The direct term is then isolated by applying a rectangular filter mask to the Fourier spectrum of the hologram, as indicated by the yellow dashed region. This selected region is later shifted to the center of the Fourier plane before performing an inverse fast Fourier transform (IFFT). The resulting complex-valued image contains both the amplitude and phase information of the object, where the transmission function is obtained from the magnitude of the reconstructed complex field, while the phase distribution is obtained from its argument. Phase unwrapping is applied to the retrieved phase image, without performing any additional background or reference subtraction.

\section{Results and discussion}

In the following, the imaging capabilities of the proposed system are demonstrated using samples with varying sizes and structural complexity. For each measurement, the translation speed of the signal mirror ($v_{s}$) is adjusted, thereby controlling the spatial frequency of the synthetic hologram fringes and the separation of the Fourier terms during reconstruction.

\subsection{Image reconstruction of binary objects}
\begin{figure}[ht!]
\centering\includegraphics[width=0.825\textwidth]{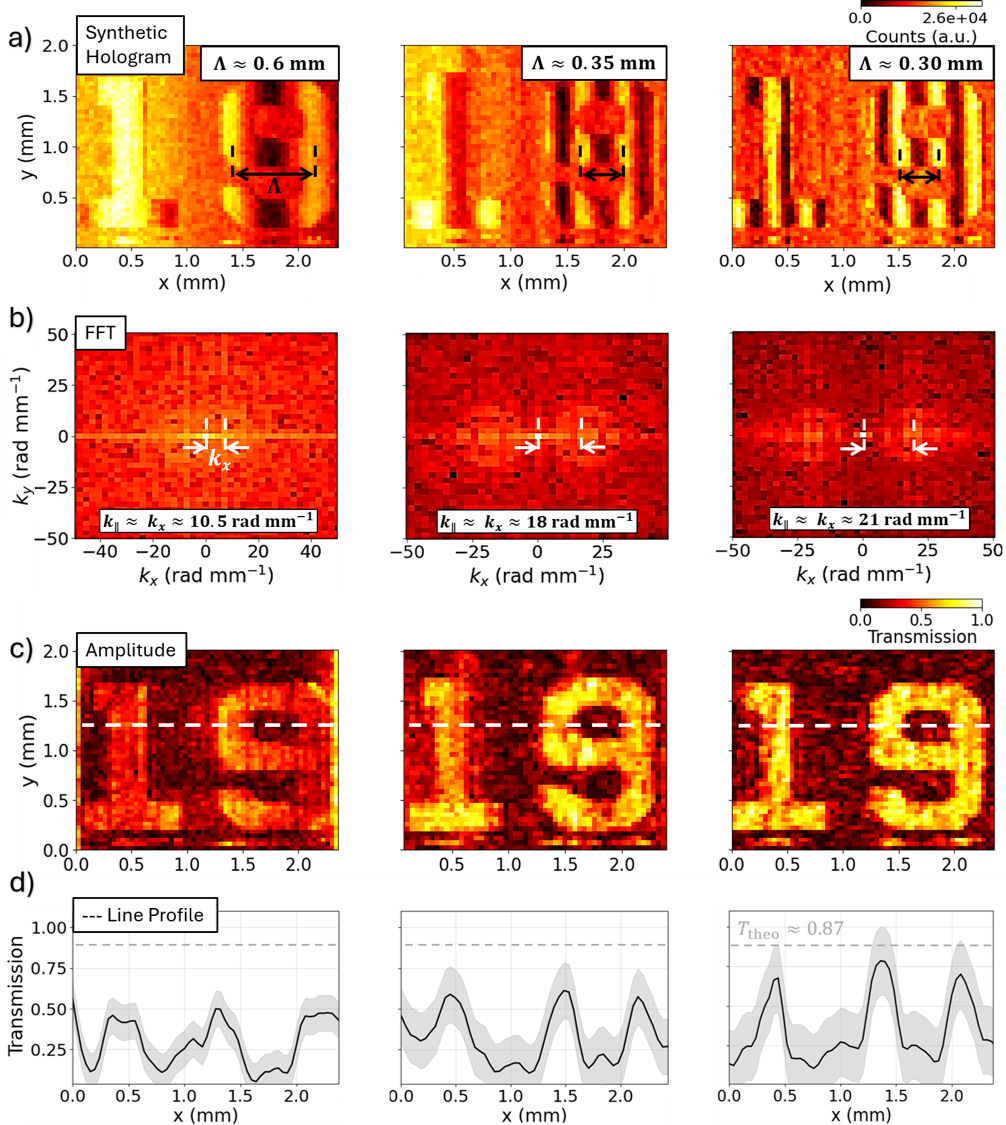}
\caption{Reconstructed amplitude and transmission profiles of a binary object featuring the number “19”, acquired with  $\Delta_x=40~\mu\mathrm{m}$ and a total acquisition time of 52 minutes. (a) Synthetic holograms recorded with three different spatial periods (0.6~mm, 0.35~mm, and 0.30~mm). (b) Corresponding FFTs images and in (c) amplitude images, where contrast variations reflect the underlying structure of the object. (d) Line profiles extracted along the white dashed lines shown in panel (c).}
\label{fig:Fig_3}
\end{figure}

An initial validation of the proposed scheme was performed using a binary mask of a USAF resolution target, where the number “19” served as the test object for evaluating the transmission function, as shown in Fig.~(\ref{fig:Fig_3}). For visualization purposes, the reconstructed images have been digitally rotated, although the original fringe orientation is horizontal. Three different fringe densities were investigated, with spatial periods of 0.6~mm, 0.35~mm, and 0.30~mm, as illustrated in Fig.~(\ref{fig:Fig_3}a).

The spatial period $\Lambda$ of the fringes is directly related to the applied phase modulation and determines the spatial separation of the interference terms in Fourier space \cite{Cuche, Sanchez}, as shown in Fig.~(\ref{fig:Fig_3}b). The corresponding amplitude reconstructions are presented in Fig.~(\ref{fig:Fig_3}c), where the effect of fringe density is observed. Increasing the fringe density enlarges the separation between the direct and conjugate terms in the Fourier domain, thereby reducing spectral overlap and facilitating a more accurate retrieval of the amplitude image \cite{Sanchez, Cuche, Leon-Torres2024}. The maximum achievable separation is ultimately limited by the detector pixel size, as the shortest resolvable fringe period must be at least twice the pixel pitch.

To quantify this effect, line profiles extracted along the white dashed lines in Fig.~(\ref{fig:Fig_3}c) are plotted in Fig.~(\ref{fig:Fig_3}d), where the theoretical transmission profile is indicated by the gray dashed line. The profiles clearly demonstrate the influence of fringe density on the reconstructed transmission function. As the fringe density increases (i.e., the spatial period decreases), the image contrast and feature definition improve. Consequently, the reconstruction with the longest spatial period exhibits the lowest contrast, whereas the reconstruction with the shortest spatial period achieves the highest contrast and best feature definition.

\subsection{Image reconstruction of transparent objects}
\begin{figure}[ht!]
\centering\includegraphics[width=0.65\textwidth]{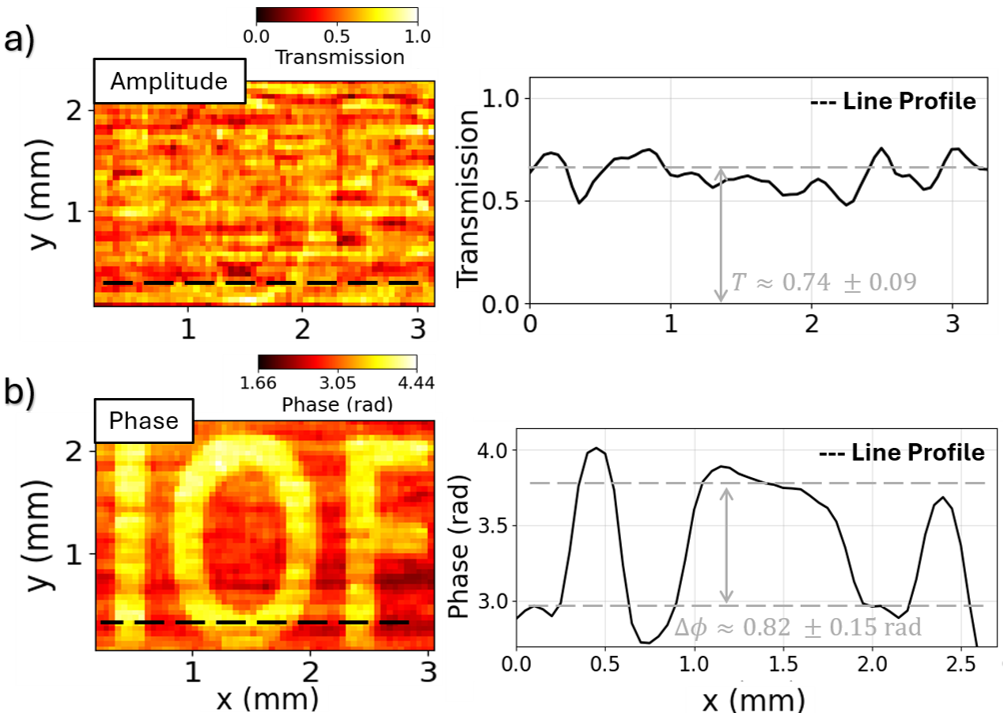}
\caption{Experimental reconstruction of the transmission amplitude and phase profiles. (a) Two-dimensional spatial distribution of the amplitude transmission (left) and its corresponding horizontal line profile (right) taken along the dashed line, yielding an average transmission of $T\approx 0.74\pm 0.09$. (b) Two-dimensional phase distribution map revealing the "IOF" structure (left) and the extracted horizontal phase line profile (right), showing a characteristic phase shift of $\Delta \phi\approx0.82\pm0.15\ \mathrm{rad}$. The image was reconstructed using a step size of $50~\mu\mathrm{m}$, resulting in a total acquisition time of approximately $55~\mathrm{min}$.}
\label{fig:Fig_4}
\end{figure}

A second measurement is performed using a transparent glass-plate with the letters "IOF" engraved on it, to assess the amplitude and phase image reconstruction of a truly transparent object, as shown in Fig.~(\ref{fig:Fig_4}). As expected the amplitude image in this case barely contains the object features, nonetheless the sharp edges of the letters "IOF" are discernible, this is caused by the finite spatial resolution of our imaging system that allows for destructive interference at the edges of the object and therefore revealing its presence, shown on the left of Fig.~(\ref{fig:Fig_4}a). A relatively flat line profile of the transmission is shown on the right of Fig.~(\ref{fig:Fig_4}a).

In contrast, the phase image reveals the object features. The "IOF" letters are easily discernible in the left panel of Fig.~(\ref{fig:Fig_4}b). A line profile extracted from the phase image yields a phase contrast of $\Delta \phi_{\mathrm{IOF}} \approx 0.82 \pm 0.15\ \mathrm{rad}$. This result agrees well with the theoretical value of $\Delta \phi_{\mathrm{theo}} = \frac{2 \pi}{\lambda_i} [2 d(n_{\mathrm{glass}} - n_{\mathrm{air}})] \approx 0.74\ \mathrm{rad}$, with the observed deviation lying within the experimental uncertainty of the measurement. Here, $d$ denotes the depth of the engraved glass features $(d_{\mathrm{IOF}}\sim250~\mathrm{nm}~ \text{and}~ d_{\mathrm{USAF}}\sim1.45~\mu \mathrm{m})$, while $n_{\mathrm{glass}}=1.57$ and $n_{\mathrm{air}}=1$ are the refractive indices for the phase objects and air at the idler wavelength, respectively.

\begin{figure}[ht!]
\centering\includegraphics[width=0.75\textwidth]{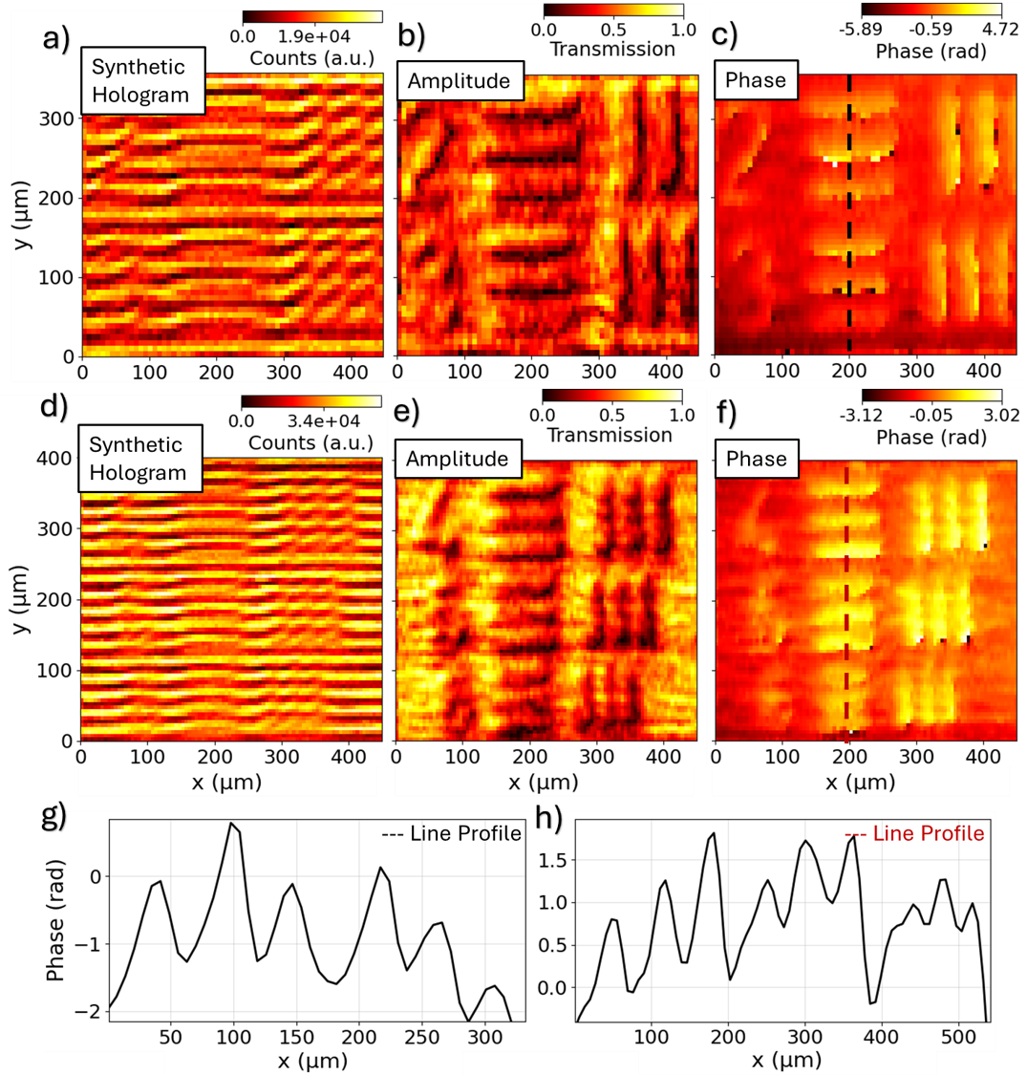}
\caption{Amplitude and phase reconstruction of the transparent USAF resolution target containing elements 2–6 of Group 4. (a,d) Synthetic holograms. (b,e) Reconstructed amplitude images. (c,f) Corresponding quantitative phase reconstructions. (g,h) Line profiles extracted along the dashed lines shown in panels (c) and (f), illustrating the phase contrast and confirming the ability of the system to resolve the target features.}
\label{fig:Fig_5}
\end{figure}

A third measurement is performed using a transparent USAF resolution target to evaluate the imaging capabilities of the system for smaller spatial features, as shown in Fig.~(\ref{fig:Fig_5}). 

Elements 2 to 6 of group 4 are illuminated. Figures~(\ref{fig:Fig_5}a–c) show the synthetic hologram, amplitude, and phase images, respectively, for elements 2 and 3, obtained with a transverse step size of 7~$\mu$m, where both the numerical features and the corresponding bar patterns are clearly resolved. Elements 4 to 6 are presented similarly in Figs.~(\ref{fig:Fig_5}d–f), acquired with a transverse step size of 5~$\mu$m. Line profiles extracted along the black and red dashed lines indicated in Figs.~(\ref{fig:Fig_5}c) and~(\ref{fig:Fig_5}f) illustrate the phase contrast across the different object features, as shown in Figs.~(\ref{fig:Fig_5}g-h). In this case, the experimentally retrieved phase contrast, $\Delta \phi_{\mathrm{USAF}} = 1.3 \pm 0.32~\mathrm{rad}$, is notably lower than the expected value of $4.18~\mathrm{rad}$. This discrepancy is attributed to the limited spatial resolution of the imaging system, as the illuminated features lie close to its resolution limit. Consequently, the high-spatial-frequency phase variations are not fully resolved. In addition, the reduced separation of the interference terms in Fourier space for these fine features results in partial overlap with the zero-order contribution, further degrading the phase reconstruction. As a result, while SOH accurately retrieves the phase of larger transparent structures, the phase contrast of features approaching the resolution limit is systematically underestimated.

Additionally, the amplitude images exhibit clear features of the object, in contrast to those shown in Fig.~(\ref{fig:Fig_4}a). As discussed previously, this behavior arises from the finite spatial resolution of the system and is particularly pronounced in this case, as the characteristic dimensions of the object are considerably smaller than those resolved in Fig.~(\ref{fig:Fig_4}a), making them comparable to the illumination beam diameter ($27.8 \pm 5\ \mu\mathrm{m}$).

\subsection{Image reconstruction of biological samples}

\begin{figure}[ht!]
\centering\includegraphics[width=0.65\textwidth]{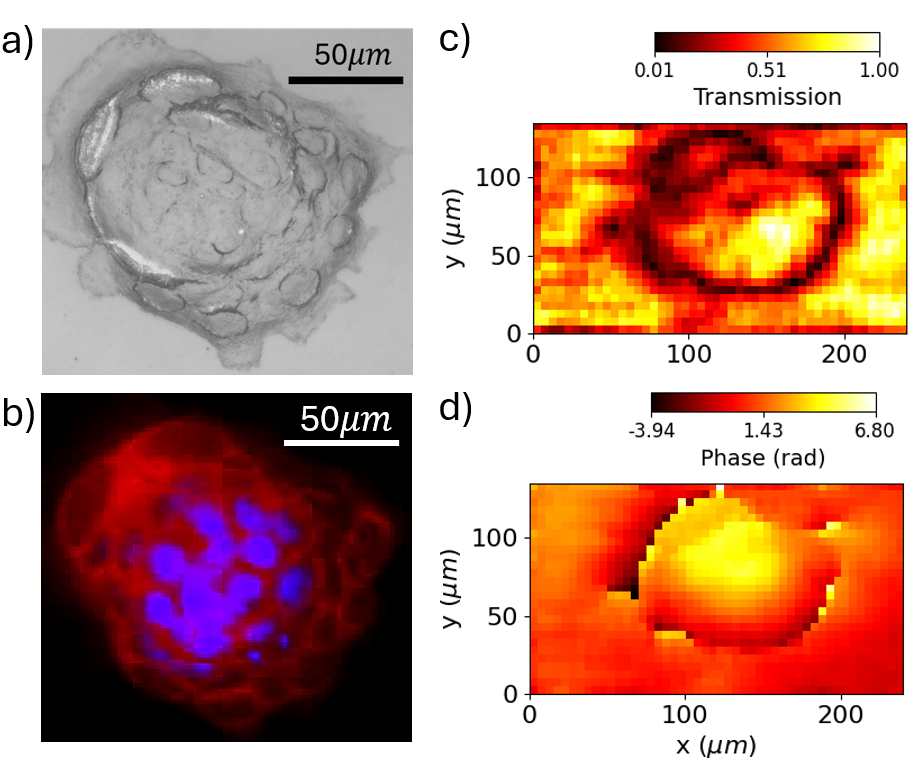}
\caption{Imaging of a cluster of FaDu cells. (a) Bright-field microscope image of the specimen. (b) Corresponding immunofluorescence image from the same region of the sample. Laminin is shown in red, while cell nuclei counterstained with DAPI are shown in blue. Scale bars are shown in each panel.
(c) Reconstructed amplitude image. The dark regions at the sample boundaries reveal increased scattering and attenuation of light at the edges of the specimen. (d) Corresponding quantitative phase reconstruction. Although neither the amplitude nor the phase image resolves the internal structure of the sample, both provide information about its overall morphology.} 
\label{fig:Fig_6}
\end{figure}

Much like the evolution of classical microscopy, where scanning techniques emerged as a powerful complement to wide-field imaging, scanning SOH based on QIUL has the potential to broaden the scope of quantum imaging applications by providing a flexible and practical alternative to existing wide-field approaches. To demonstrate these capabilities, a biological sample consisting of FaDu cells, a human pharyngeal carcinoma cell line cultured on glass coverslips and subsequently fixed with 4$\%$ paraformaldehyde (PFA), was analyzed and its transmission and phase profiles were reconstructed, as shown in Fig.~(\ref{fig:Fig_6}).

The bright-field reflection microscope image of the sample is presented in Fig.~(\ref{fig:Fig_6}a), where its overall morphology is visible and a distinct ring-shaped structure can be identified. The corresponding immunofluorescence image from the same region of the sample is shown in Fig.~(\ref{fig:Fig_6}b). Laminin, a major component of the extracellular matrix and basement membrane, was detected using an Alexa Fluor$^{\mathrm{TM}}$ 546-conjugated secondary antibody (red), while cell nuclei were counterstained with DAPI (blue). The laminin signal highlights the extracellular matrix surrounding the cells, whereas DAPI identifies the cell nuclei. Together, these fluorescence markers provide complementary biological context for the structures observed through the quantum scanning SOH scheme.

The reconstructed transmission image, shown in Fig.~(\ref{fig:Fig_6}c), reveals the ring-like morphology through scattering and attenuation at the sample boundaries. The corresponding quantitative phase reconstruction is presented in Fig.~(\ref{fig:Fig_6}d), where pronounced optical path length variations reveal the morphology of the specimen. By comparison with the immunofluorescence image in Fig.~(\ref{fig:Fig_6}b), regions exhibiting distinct phase variations can be associated with the antibody-labeled extracellular structures (orange) and cell nuclei (yellow), demonstrating the capability of SOH to retrieve label-free structural information from the sample. The finer internal structures of the specimen are not resolved, as their characteristic dimensions lie below the spatial resolution limit of the current imaging system. The images were acquired using a 5~$\mu$m scanning step, resulting in a total acquisition time of approximately 38 minutes.

\section{Conclusion}


By integrating SOH into the QIUL scanning scheme, this work further establishes a complete imaging modality capable of retrieving both the transmission and phase functions of an object using undetected light. The ability to perform full complex-field reconstruction with a single acquisition at each scan position enables efficient and robust imaging while preserving the key advantage of probing the sample in the MIR and performing detection at visible wavelengths. This combination provides a practical route toward diffraction-limited, label-free MIR imaging using mature, highly sensitive, and low-noise detection technologies.

The capabilities of the system were experimentally validated through the imaging of objects with progressively increasing structural complexity. First, a binary USAF resolution target was used to evaluate our method. Subsequently, a series of engraved glass samples containing features with dimensions ranging from several millimeters down to approximately 17~$\mu$m were analyzed, demonstrating the ability of the technique to quantitatively recover both transmission and phase information across multiple length scales. 

Finally, the method was applied to the imaging of a biological specimen. The reconstructed transmission and phase distributions revealed the overall morphology of the sample, while complementary immunofluorescence imaging provided biological context by identifying the laminin-rich extracellular matrix surrounding the cells and the cell nuclei stained with DAPI. This multimodal characterization facilitates the interpretation of the quantum imaging data. Together, these results demonstrate the versatility of the proposed approach across a broad range of imaging scenarios and highlight its potential for label-free biomedical imaging.

More broadly, the presented quantum SOH framework bridges the gap between proof-of-principle demonstrations of quantum imaging and the practical requirements of real-world imaging applications. By combining diffraction-limited spatial resolution, quantitative phase retrieval, MIR probing, and visible-wavelength detection within a scanning architecture, the technique provides a scalable platform for advanced quantum imaging and microscopy.

\begin{backmatter}
\bmsection{Funding}

The authors acknowledge support by the Carl-Zeiss-Stiftung within the Carl-Zeiss-Stiftung Center for Quantum Photonics (CZS QPhoton) under the project ID P2021-00-019, and the Deutsche Forschungsgemeinschaft (DFG, German Research Foundation) under Germany´s Excellence Strategy – EXC 2051 – Project-ID 390713860. In addition, this work was supported by the Horizon WIDERA 2021-ACCESS-03-01 grant 101079355 "BioQantSense", from the European Union’s Horizon 2020 Research and Innovation Action under Grant Agreement No. 101113901 (Qu-Test, HORIZON-CL4-2022-QUANTUM-05-SGA) and by grants funded by the Federal Ministry of Research, Technology and Space (QUANTIFISENS, 03RU1U071M and QUANCER, 13N16441).

\bmsection{Acknowledgments}
We would like to show our gratitude to R. Leitel from Fraunhofer IOF for preparing the glass samples used in the measurements.

\bmsection{Disclosures}

The authors declare no conflict of interest.

\bmsection{Data availability}

The data that support the findings of this study are available from the corresponding author, J.R.L.T., upon reasonable request.



\end{backmatter}

\bibliography{Optica-template}

\end{document}